\documentclass[a4paper]{jpconf}
\usepackage{graphicx}
\begin{document}
\title{Polarization modes for strong-field gravitational waves}

\author{Bethan Cropp and Matt Visser}

\address{School of Mathematics, Statistics, and Operations Research \\
Victoria University of Wellington \\
Wellington, New Zealand}

\ead{bethan.cropp@msor.vuw.ac.nz, matt.visser@msor.ac.nz}

\def\d{{\mathrm{d}}}
\newcommand{\scri}{\mathscr{I}}
\newcommand{\sun}{\ensuremath{\odot}}
\def\J{{\mathscr{J}}}
\def\sech{{\mathrm{sech}}}
\def\T{{\mathcal{T}}}
\def\tr{{\mathrm{tr}}}

\begin{abstract}
Strong-field gravitational plane waves are often represented in either the Rosen or Brinkmann forms. These forms are related by a coordinate transformation, so they should describe essentially the same physics, but the two forms treat polarization states quite differently. Both deal well with linear polarizations, but there is a qualitative difference in the way they deal with circular, elliptic,  and  more general polarization states. In this article we will describe a general algorithm for constructing arbitrary polarization states in the Rosen form.
\end{abstract}

\section{Brinkmann form} 

Consider the general $pp$ spacetime geometry~\cite{exact, Penrose1, Penrose2, pp, griffiths} 
\begin{equation}
\d s^2 = - 2\,  \d u \, \d v + H(u,x,y) \, \d u^2+ \d x^2 + \d y^2. 
\end{equation}
It is then a standard result that the only nonzero component of the  Ricci tensor is
\begin{equation}
R_{uu}= -{1\over2} \left\{ \partial_x^2 H(u,x,y) + \partial_y^2H(u,x,y) \right\}.
\end{equation}
Restricting attention to vacuum plane waves~\cite{Brinkmann} gives us the form:
\begin{equation}
\label{E:1}
\fl \d s^2 = - 2\,  \d u \, \d v + \left\{ [ x^2-y^2] \, H_+(u) + 2xy \, H_\times(u) \right\} \, \d u^2+ \d x^2 + \d y^2. 
\end{equation}
In this form of the metric the two polarization modes are explicitly seen to decouple. By choosing $H_+(u)$ and $H_\times(u)$ appropriately we can construct any general polarization state.
\section{Rosen form}

The ``most general''  form of the Rosen metric is~\cite{exact,Rosen}
\begin{equation}
\d s^2 = - 2 \, \d u\, \d v + g_{AB}(u)\;\d x^A\,\d x^B,
\end{equation}
where $x^A = \{ x,y\}$. 
The only non-zero component of the Ricci tensor is~\cite{exact}:
\begin{equation}
R_{uu} = - \left\{ {1\over2} \; g^{AB} \; g_{AB}'' 
- {1\over4} \; g^{AB} \; g_{BC}' \; g^{CD} \; g_{DA}' \right\}.
\end{equation}
Though relatively compact, because of the implicit matrix inversions this is a grossly nonlinear function of the metric components. In particular, in this form of the metric the $+$ and $\times$ linear polarizations do not decouple in any obvious way~\cite{Cropp}.

\section{Linear Polarization}

Consider the strong-field gravity wave metric in the $+$ linear polarization. That is, set $g_{xy} = 0$. It is found most useful to put the resulting metric in the form~\cite{Cropp}
\begin{equation}
\d s^2 = - 2\,\d u\;\d v + S^2(u) \; \left\{ e^{+X(u)} \;\d x^2 +  e^{-X(u)} \d y^2 \right\}.
\end{equation}
Then
\begin{equation}
R_{uu} = - {1\over2} \left\{ 4 \; {S''\over S} + {(X')^2}  \right\}.
\end{equation}
In vacuum we have the general vacuum wave for $+$ polarization in the form
\begin{equation}
\label{E:3}
\d s^2 = - 2\, \d u\;\d v + S^2(u) \; \left\{ \exp\left(  2 \int^u  \sqrt{-S''/S} \; \d u \right)  \;\d x^2 +  \exp\left(  -2 \int^u  \sqrt{-S''/S} \; \d u \right) \d y^2 \right\}.
\end{equation}
From the $+$ polarization, by rotating the $x$--$y$ plane through a fixed but arbitrary angle $\Theta_0$, we can easily deal with linear polarization modes along any desired axis.

\section{Arbitrary Polarization}

Now take an arbitrary, possibly $u$ dependent, polarization and consider the following metric ansatz~\cite{Cropp}:
\begin{eqnarray}
\d s^2 &=& - 2\, \d u\;\d v  + S^2(u) \; \left\{ \vphantom{\Big|} [ \cosh(X(u)) + \cos(\theta(u)) \sinh(X(u))] \d x^2 \right.
\nonumber
\\
&&
\qquad \left. \vphantom{\Big|} + 2 \sin(\theta(u)) \sinh(X(u)) \d x \, \d y   +  [\cosh(X(u)) - \cos(\theta(u)) \sinh(X(u))] \d y^2 \right\}. \;\; 
\end{eqnarray}
Note setting $\theta(u) = \Theta_0$ corresponds to linear polarization. The vacuum field equations imply
\begin{equation}
  4 \, {S''\over S} + {(X')^2}  + \sinh^2(X(u)) \;  (\theta')^2 = 0.
\end{equation}
Let us introduce a dummy function $L(u)$ and split this into the two equations
\begin{equation}
 4\, {S''\over S} + {(L')^2}   = 0, \qquad (L')^2 =  {(X')^2}  + \sinh^2(X(u)) \;  (\theta')^2.
\end{equation}
The first of these equations is just the equation you would have to solve for a pure $+$ (or in fact any linear) polarization.
The second of these equations can be rewritten as
\begin{equation}
\d L^2 =  \d X^2  + \sinh^2(X) \;  \d\theta^2,
\end{equation}
and is the statement that $L$ can be interpreted as distance in the hyperbolic plane $H_2$.

Compare this to Maxwell electromagnetism, where polarizations can be specified by 
\begin{equation}
\vec E(u)  = E_x(u) \; \hat x + E_y(u)\; \hat y,
\end{equation}
with no additional constraints. Thus an electromagnetic wavepacket of arbitrary polarization can be viewed as an arbitrary ``walk'' in the $(E_x, E_y)$ plane. We could also go to a magnitude-phase representation $(E,\theta)$ where
\begin{equation}
\vec E(u)  = E(u) \cos\theta(u) \; \hat x + E(u) \sin\theta(u)\; \hat y.
\end{equation}
So an electromagnetic wavepacket of arbitrary polarization can also be viewed as an arbitrary ``walk'' in the $(E, \theta)$ plane, where the $(E, \theta)$ plane is provided with the natural Euclidean metric
\begin{equation}
\d L^2 = \d E^2 + E^2 \; \d \theta^2.
\end{equation}
In contrast for gravitational waves in the Rosen form we  are now dealing with an arbitrary ``walk'' in the hyperbolic plane, $H_2$. Furthermore, because of the nonlinearity of general relativity, there is still one remaining  differential equation to solve.

\section{Circular Polarization}

As an important example, we consider strong-field circular polarization. Circular polarization corresponds to a fixed distortion $X_0$ with a linear advancement of $\theta(u)$:
\begin{equation}
\theta(u) = \Omega_0\; u; \qquad   X(u) = X_0.
\end{equation}
Then~\cite{Cropp}
\begin{eqnarray}
\d s^2 &=& - 2\, \d u\;\d v  + S^2(u) \; \left\{  \vphantom{\Big|} [ \cosh(X_0) + \cos(\Omega_0\; u) \sinh(X_0)] \d x^2 \right.
\nonumber
\\
&&
\qquad \left.  \vphantom{\Big|} + 2 \sin(\Omega_0 u) \sinh(X_0) \d x \, \d y   +  [\cosh(X_0) - \cos(\Omega_0 u) \sinh(X_0)] \d y^2 \right\}. 
\end{eqnarray}
So the only nontrivial component of the Ricci tensor is
\begin{equation}
R_{uu} = - {1\over2} \left\{ 4 \; {S''\over S}   + \sinh^2(X_0) \;  \Omega_0^2 \right\}.
\end{equation}
Solving the vacuum equations gives
\begin{equation}
S(u) = S_0 \; \cos\left\{ { \sinh(X_0) \;  \Omega_0 \; (u-u_0) \over 2}\right\}.
\end{equation}
This now describes a spacetime that has good reason to be called a strong-field circularly polarized gravity wave. Note the weak-field limit corresponds to $X_0\ll 1$ so for an arbitrarily long interval in $u$ we have  $S \approx S_0$, and without loss of generality we can set $S\approx 1$. Then, as expected, we obtain~\cite{Cropp}
\begin{equation}
\d s^2 \approx  - 2\, \d u\;\d v  + \d x^2 + \d y^2 + X_0 \; \left\{  \vphantom{\Big|} \cos(\Omega_0\; u) [\d x^2-\d y^2] + 2   \sin(\Omega_0\; u) \;\d x \;\d y \right\}.
\end{equation}

\section{Decoupling the general form}

Let us return to considering the metric in general Rosen form
\begin{equation}
\d s^2 = - 2\, \d u\; \d v + g_{AB}(u)\;\d x^A\,\d x^B, 
\end{equation}
where  $x^A$, $x^B$  represent any arbitrary number of dimensions ($d_\perp\geq 2$) transverse to the $(u,v)$ plane. 
It is easy to check that the only non-zero component of the Ricci tensor is still
\begin{equation}
R_{uu} = - \left\{ {1\over2} \; g^{AB} \; g_{AB}'' 
- {1\over4} \; g^{AB} \; g_{BC}' \; g^{CD} \; g_{DA}' \right\}.
\end{equation}
Let us now decompose the $d_\perp \times d_\perp$ matrix $g_{AB}$ into an ``envelope'' $S(u)$ and a unit determinant related to the ``direction of oscillation''~\cite{Cropp}. That is, let us take
\begin{equation}
g_{AB}(u) = S^2(u)\; \hat g_{AB}(u),
\end{equation}
where $\det( \hat g) \equiv 1$. (A related discussion can be found in~\cite{Landau}.) Calculating the various terms of the Ricci tensor using the relations 
\begin{equation}
  [  \hat g^{AB} \; \hat g_{AB}']  = 0, \qquad  [  \hat g^{AB} \; \hat g_{AB}''] -  [  \hat g^{AB} \; \hat g_{BC}' \; \hat g^{CD} \; \hat g_{DA}'  ] = 0,
  \end{equation}
it is found that~\cite{Cropp}
\begin{equation}
R_{uu} =  - d_\perp \,  {S''\over S}   - {1\over2}\, [  \hat g^{AB} \; \hat g_{BC}' \; \hat g^{CD} \; \hat g_{DA}'  ].
\end{equation}
Note that we have now succeeded in decoupling  the determinant ($\det(g) = S^{2d_\perp}$; effectively the  ``envelope'' $S(u)$ of the gravitational wave) from the unit-determinant matrix $\hat g(u)$.

Now consider the set $SS(I\!\!R, d_\perp)$ of all unit determinant real symmetric matrices, and on that set consider the Riemannian metric
 \begin{equation}
 \d L^2 =  \tr\left\{  [\hat g]^{-1} \;\d[\hat g] \; [\hat g]^{-1} \; \d[\hat g]  \right\}.
\end{equation}
Then
\begin{equation}
R_{uu} =  -  {1\over2} \left\{ 2 d_\perp\, {S''(u)\over S(u)}   + \left({\d L\over\d u}\right)^2  \right\}. 
\end{equation}
This means an arbitrary polarization vacuum Rosen wave is an arbitrary walk in $SS(I\!\!R, d_\perp)$,  with distance along the walk $L(u)$ related to the envelope function  $S(u)$ as in the discussion above. 

\section{Discussion}

Arbitrary polarizations, while trivial in the Brinkmann form, are difficult to implement in the Rosen form.
To address this  puzzle we have re-analyzed the Rosen strong-field gravity wave in terms of an ``envelope'' function and two freely specifiable functions. The vacuum field equations can be interpreted in terms of a single differential equation governing the ``envelope'', coupled with an arbitrary walk in polarization space.  In particular we have indicated how to construct a circularly polarized Rosen form gravity wave,  and how to generalize this central idea beyond (3+1) dimensions. Further detailed calculations and discussions of these Rosen form polarizations can be found in \cite{Cropp}.

\ack

This research was supported by the Marsden Fund administered by the Royal Society of New Zealand.

\end{document}